# Introduction to the African Strategy for Fundamental and Applied Physics (ASFAP)

Farida Fassi

*Mohammed V University in Rabat, Faculty of Science, Morocco*

**Abstract:**
Generating scientific and technological knowledge and converting them into innovations which are of added value to society are key instruments for a society's economic growth and development. As outstanding as these capabilities are for other regions in the world, Africa's science, innovation, education and research infrastructure, particularly in fundamental and applied physics, have over the years been under-valued and under-resourced. To efficiently address the scientific and technological gaps with the rest of the world, Africa's stance needs radical overhaul. With the big ambition to drive a community-wide effort in Africa, the African Strategy for Fundamental and Applied Physics (ASFAP) was founded. The aspiration is to demonstrate the physics potential benefits for African society and how physics can contribute to the technological infrastructure development and to provide trained personnel needed to take advantage of scientific advances. The vision consists in fostering scientific literacy driven by physics-based technologies and their impact for economic growth, including other sciences that draw heavily on advances in physics. In addition to developing and enhancing collaborations and partnerships among Africans in national, regional, and Pan-African organizations.  This should assist to tackle the challenges that Africans struggle and prioritize educational and research resources, innovation and development. The ASFAP initiative could present a unique opportunity of overcoming the complexity of the African social and economic challenges, if Africa needs to have and maintain its position as a co-leader in the global scientific process and reap the consequent socio-economic benefits. ASFAP will take a few years with a final report to notify the African policymakers and broader communities concerning the strategic directions that will have greatest impacts on physics education and research in the next decade.

## 1. Introduction

Science and technology capabilities are fundamental for social and economic growth and development. Yet Africa's science, innovation and education have been chronically under-funded. Transferring knowledge, building research capacity and developing competencies through training and education are major priorities for Africa in the 21st century. Physics combines these priorities by extending the frontiers of knowledge and inspiring young people. It is therefore essential to make basic knowledge of emerging technologies available and accessible to all African citizens to build a steady supply of trained and competent researchers. In this spirit, ASPAF [1] was founded to foster social transformation and economic competitiveness, through science, technology and innovation

Corresponding Author
Email address:farida.fassi@cern.ch (Farida Fassi)

for effective human capital development as a key means of implementation to drive sustainable development in Africa. ASFAP aims to increase African education and research capabilities, build the foundations and frameworks to attract the participation of African physicists, and establish a culture of awareness of grassroots physics activities contrary to the top-down strategies initiated by governments. To address many of the fundamental issues that African society continue to face, ASFAP will complement African top-down strategies and encourage a broad community participation.

**2. Africa Major Challenges in the 21th century**

Literacy and education are essential for human development in today's knowledge world as Nelson Mandela said "*Education is the most powerful weapon, which we can use to change the world*". Education allows us to better understand the world in which we live. Nevertheless, literacy in the African continent is affected by a number of key factors, which are: colonial legacy includes the linguistic framework that strongly affect the educational environment, as well as shortage of resources (reliable access to electricity), inadequate infrastructure as shown in Graph-1 (middle and right-hand images), lack of digital literacy, lack of trained teachers and better learning material which include Information and Communication Technologies (ICT). These constraints hinder the process of promoting literacy to Africans, and present the biggest challenges of science knowledge creation and technological innovation in Africa. These led to the continent´s failure to create high quality education, stop brain drain, correct for gender equality, as well as they fail to ensure science-led development on a sustainable basis and unlock the minds for brighter economic prospects.

Because of that, the overwhelming majority of the Africans are falling into poverty and lack of decent jobs. Parallel to this, there is a demographic explosion in Africa (see Graph-1, left-hand) based on the United Nations forecasts [2] [3]. Africa's population is expected to grow by an average of 2.2 % (2.5 % by 2040) every year until the 2060s (and by 2050 a quarter of the world's people will be African). In addition, more than half of Africa's population will be under 25 years old [4]. In the next few decades, Africa will become the youngest and most populous continent. In view of these developments, the difficulty of confronting a population explosion and weak economic growth exacerbated the situation. The situation for female youth, who are even worse off than male youth, is particularly serious. Young women suffer from a deeper lack of insufficient levels of literacy and therefore are largely emarginated and work in the informal and precarious sectors.

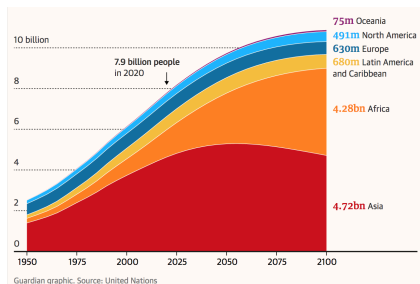 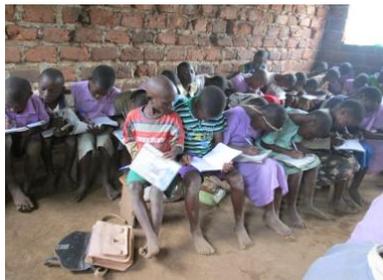 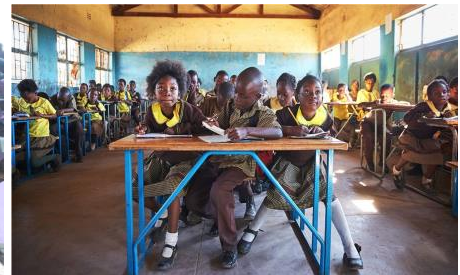

Graph-1: The world´s population will rise to 10,9 billion by 2100, with most of the growth driven by Africa [3] (Left-hand). Middle and right-hand images show examples of the educational environment issues in Africa.

Although, African countries show commitment to invest in science through investment in research and development (R&D), only four out of five countries still spent less than 1% of global investment in R&D of its Gross Domestic Product, which is the closest to the African Union's suggested of 1% [5]. Furthermore, several countries ranked as 0%, while the R&D expenditures for more than 20 African countries are unknown. This chronic underfunding on Africa's R&D is gravely impacting the Africa´s participation in the creation of scientific knowledge, which is extremely poor in comparison with the rest of the world. The Africa's contribution to the global scientific knowledge is roughly 0.74%. On top of this, African publishing research outputs are under-valued and under-cited in the international indexing systems namely, Scopus and the Web of Science. That said, Africa has also an extremely low number of researchers per million people (it is about 198) [6], and in order to achieve the world average concerning the number of researchers per capita, Africa needs another million new PhDs students. This landscape gets worse and worse when Africa is losing roughly 20,000 professionals to the high-income countries every year due to the lack of opportunity. Hence, one must recognise that the aforementioned limitations leave Africa lagging behind and the continent become a peripheral region in the world. Even, the gap is widening exacerbated by the persist underperformance. It is high time, that Africa narrows and closes the scientific and technological gaps with other regions in the world. That it's been massively underfunded over a long time impacting critically the growth of African economies. Thus, African countries are at a critical juncture socioeconomic, and to achieve the commitment of the leadership a deep African engagement is required. It is therefore, imperative that Policymakers and stakeholders appreciate the return on investment in research, innovation and higher education to reverse the trend towards greater science-led development on a sustainable basis for Africans.

## 3. ASFAP-African Strategy for Fundamental and Applied Physics

In response to the previously mentioned challenges, ASFAP presents an ambitious initiative that seeks to set up the foundation and framework for enhancing the African science community collaboration in defining education, physics priorities and stressing the powerful of physics for African society. The main goal is to address the interests of a wide range of varying needs from the community with the aim to attract enthusiasms of every physics disciplinary group, which is particularly targeted. These perspectives have arisen from and been driven by the science community with its potential to deliver a long-term strategy that would encourage and engage the continent to align its scientific research policy in physics to support national agendas towards higher education development and improve basic physics research by using more robust means of assessment that focus mainly on values of insight, impact and reliability. All these include a strong desire for investment in African science for economic growth driven by physics-based technologies.

ASFAP founding initiative is designed to be a transparent and democratic process, to be owned by Africans for Africa and it is mandated by the African Physical Society. The co-founders who form the Steering Committee (STC) are: Kétévi Assamagan (Brookhaven National laboratory, USA), Simon Connell (University of Johannesburg, South Africa), Farida Fassi (Mohammed V University in Rabat, Morocco), Shaaban Khalil (Center for Fundamental Physics, Zewail City, Egypt) and Fairouz Malek (CNRS and Grenoble University, France).

Corresponding Author  
Email address:farida.fassi@cern.ch (Farida Fassi)

The process development requires a few years and in due course will release the strategy report. STC will first explore manners to better adapt the overall ASFAP strategy generated. In particular, the focus will be suggesting a strategic plan with direction that has a positive and significant impact on physics education and research in the next decade, along with actionable items. The report will be communicated to the African and broader communities. With the aim to make well-informed decisions that would contribute to a sustainable development, a collaborative culture must be instilled as a core principal that can bring together policymakers, managers, scientific communities in national, regional and Pan-African organizations. Moreover, to strengthen this collaboration and ensure the implementation of those decisions one needs to do a close follow up.

Just like a strategic plan, in order to be useful it needs to be periodically reviewed and revised to adjust priorities and re-evaluate goals, so that ASFAP process has to be repeated regularly, every 7-10 years for the following decades with a review of the impact of previous strategies. The overall roadmap process is shown in Fig-1, that illustrates a tentative timeline and deliverables, as well as the activities.

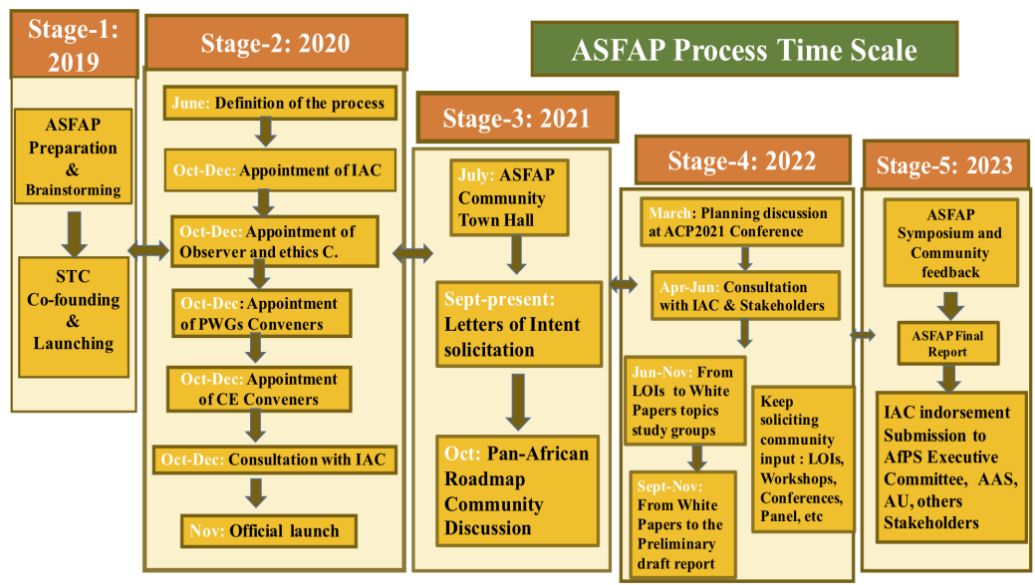

Fig-1: ASFAP roadmap timeline

ASFAP organization structure is shown in Fig-2. The Steering Committee (STC) body manages the process development, ensure the overall coordination and advance the necessary efforts involving all ASFAP bodies. In addition to the final report preparation. The International Advisory Committee (IAC) is formed of international institutes and leading individuals in the field in Africa and beyond. IAC advises on the ASFAP scope, review the progress and engage the international communities and policymakers as well as endorse the final report. ASFAP considers the Observers Committee (OC) who is made up of Seniors and Expects from many research fields in physics. OC is responsible for advising and conveying ideas between STC, Physics Working Groups (PWGs), Fora and Community Engagement groups. OC has also to review the community inputs, in particular Letter Of Intents (LOIs) and White paper, and help PWGs in report editing. ASFAP include also the Ethics

Corresponding Author 4
Email address:farida.fassi@cern.ch (Farida Fassi)

Committee (EC), who is responsible for dissemination and maintain the Guidelines of ASFAP Code of conduct. A subset of EC might serve as an Ombudspersons as well.

ASFAP has a broad footprint by discipline and fields, it has 16 PWGs and are as follows: Accelerators, Astrophysics & Cosmology, Atomic & Molecular Physics, Biophysics, Computing & 4IR, Earth Science, Energy, Fluid and Plasma, Instrumentation & Detectors, Light Sources, Materials Physics, Medical Physics, Nuclear Physics, Particle Physics, Optics and Photonics and Theoretical & Applied Mechanics. PWGs play a pivotal role in the ASFAP process, by encouraging and soliciting the community inputs, engaging the discussion, reviewing and revising the progress in the relevant groups. Furthermore, PWGs are in charge of the group reports preparation. PWGs provide also the liaisons between the different PWGs to enable that cross-cutting topics receive the proper coverage and consideration in all the relevant groups. In what concern the Engagement Groups & Fora, there are Community Engagement, Physics Education, Women in Physics Forum and Young Physicists Forum. These are convened and working closely with STC and the rest of ASFAP bodies to promote science-led development on a sustainable basis, contribute to the scientific literacy of high schools and universities and revitalize higher education. It is clear that the first step should be the gathering of inputs and evidence from the community to draw upon. This kind of inputs give the full picture and help to address the right questions on how to encourage and engage young people in physics, technology and scientific careers. And a special attention must be paid to the women participation. This widely and deeply consultation can contribute in raising awareness of the necessity to prepare and form qualified young people, to technology and knowledge transfers and can thus stimulate Africa's productivity development.

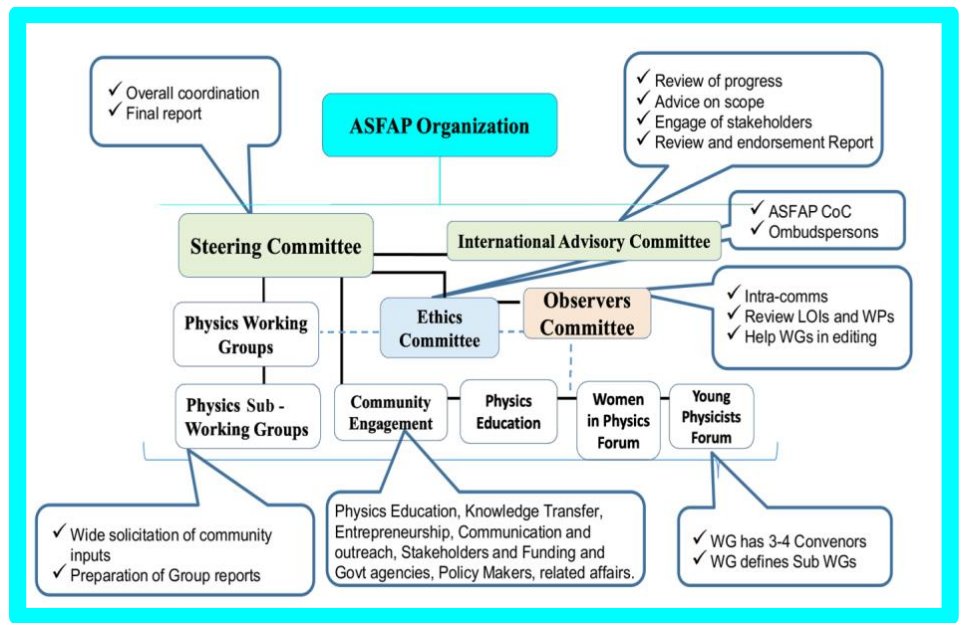

Fig-2: ASFAP Organization structure

## 4. African Strategies and ASFAP

Africa has already numerous strategies authored by governments and policymakers, large political bodies and institutions [5] [7] [8]. These include strategies for a green growth model, better education and research, job creation, digital transformation and so on. However, the effective local implementation of these strategies is still missing, and which is evident by the persistent of Africa laggardness. As a result, it is essential for the scientist's community, engineers, technicians, funding agencies and policymakers to come together and create the critical masses needed to address and define a concerted strategy for the continent. Unlike African top-down strategies, ASFAP is based on a large community consultations and large footprints in all the physics and engagement topics of importance to Africa.

Since decades other regions in the world have already been conducting the efforts; this is the case for the European and United States strategies for Particle Physics and other physics fields, and recently Latin America that has been developing its strategy as well. For that, Africa should fight seriously to take its equal place as a co-leader in the global scientific process, along with all the consequent socio-economic benefits. To achieve that, ASFAP is a crucial process, where the central mission would be to help improving higher education in Africa across national borders and in so doing, to contribute in a significant way to the development on the continent. To fulfil this mission several considerations, have to be implemented among them: 1.) the engagement and participation of the African scientist's community, 2.) the establishment of culture awareness for the periodic strategies done by grassroots physicists, 3.) the improvement of the low level of intra-African collaboration in the exchange and sharing of data and in scientific collaboration and 4.) the strengthen of the African education and research capabilities. Since the international cooperation forms the common denominator of today's culture of scientific activities, it is equally important, to increase and sustain networking to extend the existing international scientific ties to Africa, in the development of the strategical visions for fundamental and applied physics. Such as engagement in physics education, communication and outreach, toward developing countries, should be sustained also in targeted programs toward Africa. In order to exploit and overcome a realistic path for development, African countries must adopt a coherent coordinated strategy and implement what it needs to be changed to radically give science and technology their due weight in the development process. This will certainly be a unique opportunity to prosper for a continent of more than a billion people with large unmet needs but vast human potential.

## 5. Organizations supporting ASFAP

Graph-2 represents the ASFAP support, including the national and international organizations funding agencies and governments organizations, institutions and academies in Africa and beyond. The organizations supporting ASFAP are as follows: African Physical Society, African Union, African Academy of Sciences, Network of African Science Academies (NASAC), South African Department of Science and Innovation, iThemba Labs (South Africa), Hassan II Academy of Science and Technology (Morocco), Ministry of Higher Education and Scientific Research (Tunisia), South African Institute of Physics (SAIP), United Nations Educational, Scientific and Cultural Organization (UNESCO), East African Institute for Fundamental Research (UNESCO-EAIFR), Third world Academy of Sciences (UNESCO-TWAS), International Union of Pure and Applied

Physics (IUPAP), CERN (European Organization for Nuclear Research), International Centre for Theoretical Physics (ICTP), International Atomic Energy Agency (IAEA), European Physical Society (EPS), German Physical Society (DPG), Institute of Physics (IOP), French Physical Society, European Astronomical Society (EAS), Spanish Royal Physics Society (RSEF), Spanish Astronomical Society (SEA), Netherlands' Physical Society (NNV), Physical Society of Japan (JPS), Association of Asia Pacific Physical Societies (AAPPS) and Islamic World Educational, Scientific and Cultural Organization (ISESCO).

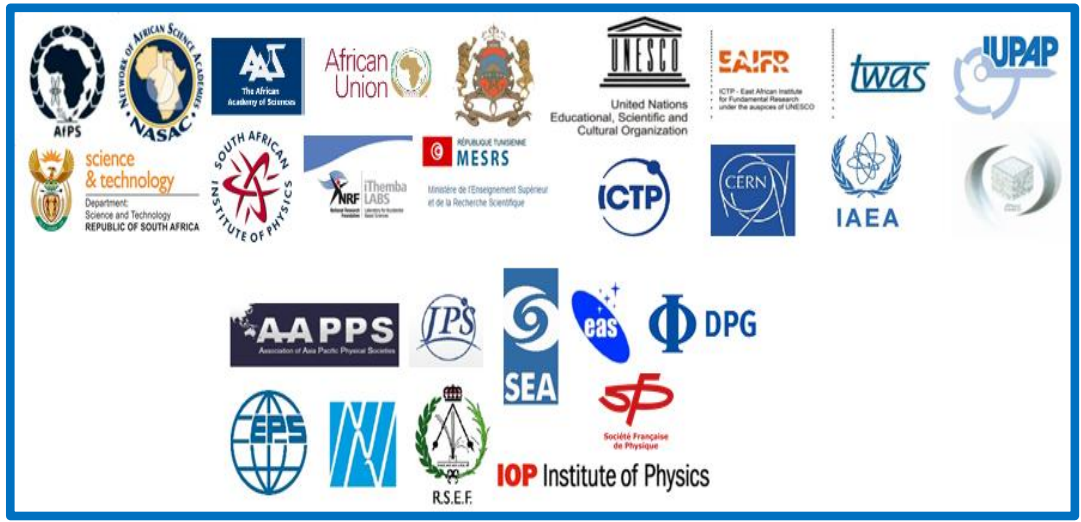

Graph-2

## 6. Community Inputs

The worldwide expert community has demonstrated the wish to promote better education and research for Africa. However, they do not consider the African conditions due to their ignorance of the African reality. To translate the diversity and complexity of African realities, Africans themselves must develop their own strategy for science and technology to achieve their greatest impact and benefits. This would also help the international partners interested in capacity development and retention in Africa to integrate the Africans inputs, rather than to default to their own views of how they may want to develop Africa. In this spirit, ASFAP process relies on the African community inputs, as well as Pan-African and international communities. The inputs from the communities are collected in several forms of proposals: LOIs, white papers (that aim to be published in a peer-reviewed Journal), including surveys, Panel discussion, networking with stakeholders, conferences and workshops and then discussed and debated in plenary sessions and topical parallel sessions. The engagement of the community inputs would be also the commitment in PWGs meetings, nominations of the PWGs conveners and liaisons, offering constructive opinions and expertise and encouraging institutional credits for those working on ASFAP for people in leadership roles as conveners and liaisons.

## 7. Conclusion

ASFAP was founded with a big ambition for change and longer-term reforms that should focus on strengthening institutions, improving infrastructure, accelerating technology adoption and enhancing high education, capacity building, scientific research. In addition, the strategy aims to influencing

Corresponding Author
Email address:farida.fassi@cern.ch (Farida Fassi)

directions of strategic science development taken by policymakers and effective implementation. ASFAP would help donors and funding agency in deciding where best to invest limited resources and support the African Physical Society into an outstanding professional body. The initiative also stresses the pivotal role that physics plays as a key part of the educational system and hence an advanced society governments need to be supportive of the science of physics. It is worth emphasizing, furthermore, that the digital revolution, the so-called 4th Industrial Revolution, presents a great opportunity for Africa to benefit from developed ICT capabilities such as automation and digitalisation if Africa succeeds in implementing the technological transformation locally.

In pursuing this vision, the African scientific communities emphasize the importance of building synergy between fundamental physics and practical applications which is crucial for a solid education in Africa.

Corresponding Author                                                                                         8
Email address:farida.fassi@cern.ch (Farida Fassi)